\documentclass{article}

\usepackage{spconf}
\usepackage{hyperref}       %
\usepackage{url}            %
\usepackage{booktabs}       %
\usepackage{nicefrac}       %
\usepackage{microtype}      %
\usepackage{amsfonts}       %
\usepackage{amsmath}
\usepackage{tabularx}
\usepackage{float}
\usepackage{outlines}
\usepackage{subcaption}
\usepackage{algorithm}
\usepackage{algpseudocode}
\usepackage{chngcntr}
\usepackage{graphicx}
\usepackage{xcolor}
\usepackage{setspace}
\usepackage{enumitem}
\usepackage{bbm}
\usepackage{pgfplots}%
\pgfplotsset{compat=1.13}%
\usepackage[htt]{hyphenat}
\usepackage{lipsum}
\usepackage{xspace}
\usepackage{multirow}
\usepackage[framemethod=TikZ]{mdframed}
\usepackage{tabu}
\usepackage{pifont}%
\usepackage{makecell}
\usepackage{cleveref}

\usepackage[square,sort,comma,numbers]{natbib}

\hypersetup{
    colorlinks,
    linkcolor={red!50!black},
    citecolor={blue!80!black},
    urlcolor={blue!80!black}
}

\renewcommand{\paragraph}[1]{\noindent\textbf{#1}\quad}

\DeclareMathSizes{10}{9}{6}{5}

\setlist{nosep} %
\setlist{itemsep=1pt, topsep=3pt}

\title{Anim-400K: A Large-Scale Dataset for Automated End to End Dubbing of Video}
\name{Kevin Cai\textsuperscript{\rm 1} \qquad Chonghua Liu\textsuperscript{\rm 1}\qquad David M. Chan\textsuperscript{\rm 1}\sthanks{Corresponding author: \url{davidchan@berkeley.edu}} }
\address{\textsuperscript{\rm 1}University of California, Berkeley}
\begin{document}
\maketitle
\begin{abstract}
The Internet's wealth of content, with up to 60\% published in English, starkly contrasts the global population, where only 18.8\% are English speakers, and just 5.1\% consider it their native language, leading to disparities in online information access. Unfortunately, automated processes for dubbing of video -- replacing the audio track of a video with a translated alternative -- remain a complex and challenging task due to pipelines, necessitating precise timing, facial movement synchronization, and prosody matching. While end-to-end dubbing offers a solution, data scarcity continues to impede the progress of both end-to-end and pipeline-based methods. In this work, we introduce Anim-400K, a comprehensive dataset of over 425K aligned animated video segments in Japanese and English supporting various video-related tasks, including automated dubbing, simultaneous translation, guided video summarization, and genre/theme/style classification. Our dataset is made publicly available for research purposes at \url{https://github.com/davidmchan/Anim400K}. 
\end{abstract}

\section{Introduction \& Background}
\label{sec:intro}

\begin{table*}
\footnotesize
\centering
    \begin{tabularx}{\linewidth}{Xccccccc}
        \toprule
        \textbf{Dataset} & \textbf{Hours} & \textbf{Clips} & \textbf{Languages} & \textbf{Source} & \textbf{Target} & \textbf{\makecell{Video \\ (Source/Target)}} \\
        \midrule
        IWSLT 2023 \cite{agrawal2023findings} & 5 & 200 & DE $\to$ EN & DE Text & EN Translation & $\checkmark$/$\times$ \\
        MuST-C \cite{di2019must} & $>385$ & $>211$K & X (8) $\to$ EN & Spoken Audio & Subtitles & $\times$/$\times$ & \\
        MSLT \cite{federmann2016microsoft} & 4 & 3K & FR/DE $\to$ EN & Audio & Translations & $\times$/$\times$ \\
        MuST-Cinema \cite{karakanta2020must} & $>385$ & $>211$K & X (7) $\to$ EN & Spoken Audio & Subtitles & $\checkmark$/$\times$ \\
        Heroes \cite{oktem2018bilingual} & 5 & 7K & ES $\leftrightarrow$ EN & Spoken Audio & Spoken Audio & $\checkmark$/$\checkmark$ \\
        \midrule
        Anim-400K (Ours) & 763 & 425K & JP $\leftrightarrow$ EN & Spoken Audio & \makecell{Spoken Audio + \\ Subtitles} & $\checkmark$/$\checkmark$ \\
        \bottomrule
    \end{tabularx}
    \caption{Overview of datasets related to automated dubbing.}
    \label{tab:related}
\end{table*}

Significant portions of the internet (up to 60\% \cite{yang2020large}) is published in English, however it is estimated that only 18.8\% of people in the world speak English, and only 5.1\% speak English as a first language \cite{cia2023world}. This language barrier can create inequities in access to information available on the web, making large amounts of high-quality information unavailable to numerous users. Much of this information is in the form of video sources, which are traditionally made accessible in one of two ways: subtitling or dubbing. In subtitling, translated subtitles are made available in a target language. In dubbing, audio tracks are replaced with audio tracks in the users' native languages. Significant research \cite{koolstra2002pros,wissmath2009dubbing,boonyubol2022comparing} has shown that dubbed videos can increasing feelings of spatial presence, transportation, and flow leading to increases in user engagement and retention. Further, dubbing makes content accessible for those who are illiterate, or those who are beginning readers.

Unfortunately, while automated subtitling has been made possible through advances in automatic speech recognition (ASR) and machine translation (MT),  dubbing translation remains a time consuming and expensive process, largely only accomplished through manual means. Recent systems for automated dubbing are based on complex pipelines, stitching together ASR, MT, and Text to Speech (TTS) systems \cite{wu2023videodubber,oktem19_interspeech}, and while advances have been made, these systems still lack complex nuance required for dubbing, including matching the timing \cite{effendi2022duration,lakew2021machine,lakew2022isometric,tam2021isochrony}, facial movements \cite{yang2020large} and the prosody \cite{oktem19_interspeech, virkar2021improvements} of the generated speech to the video. ``End-to-end dubbing", where translated audio is produced directly from raw source audio, is a potential solution to this complexity, and has numerous other benefits including the ability for the model to capture small variations in the speaker performance, a key quality of a good dub \cite{brannon2023dubbing, yang2017end}.

\begin{figure}
    \centering
    \includegraphics[width=\linewidth]{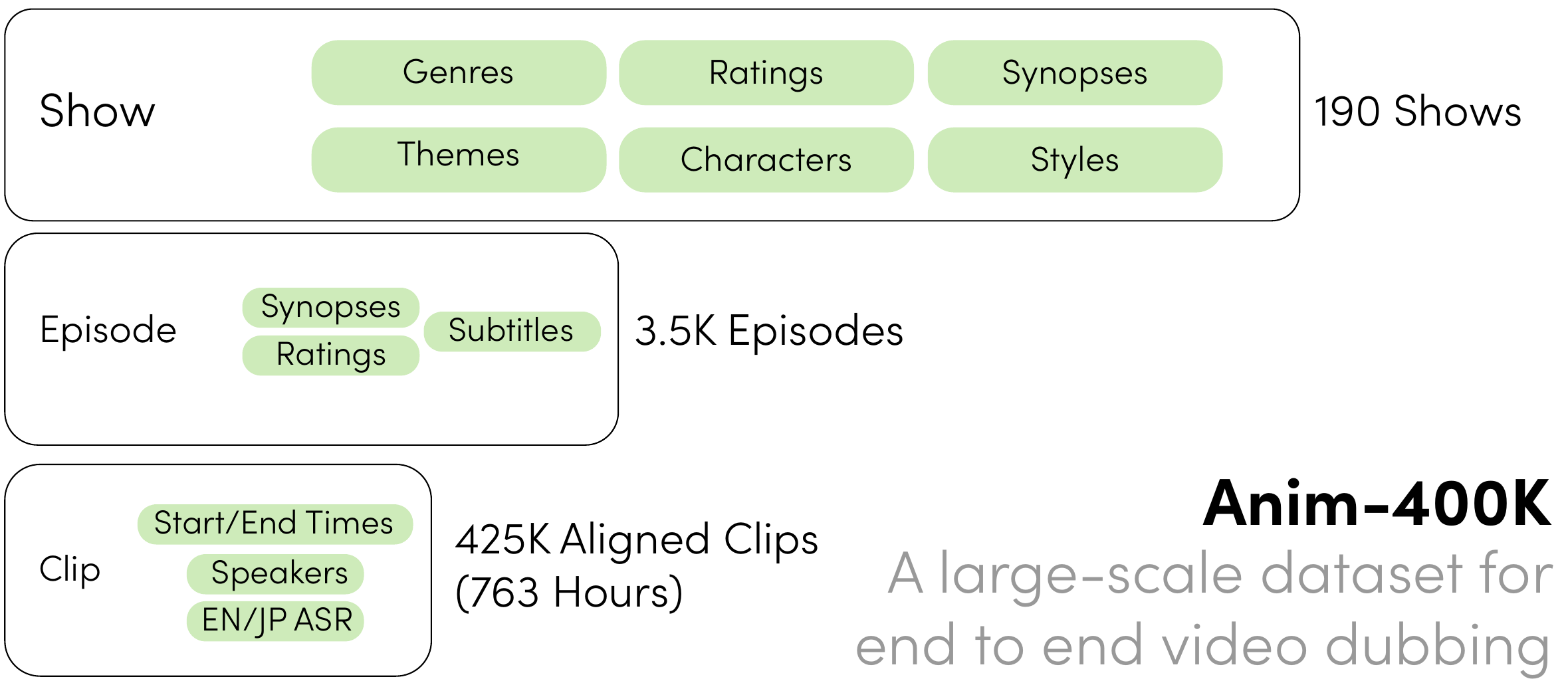}
    \caption{Anim-400K is a new dataset designed from the ground up for automated dubbing of video, and supporting a wide range of secondary video tasks ranging from simultaneous translation and guided video
summarization to genre/theme/style classification.}
    \label{fig:enter-label}
\end{figure}

Unfortunately, while end-to-end dubbing of videos is an intriguing task, there is almost no data support for the task, leading to practical limitations on the quality of end-to-end dubbing models \cite{agrawal2023findings,swiatkowski2023cross,wu2023videodubber,federico2020evaluating,lakew2021machine,lakew2022isometric,tam2021isochrony,effendi2022duration}. Almost all prior work identifies the Heroes corpus \cite{oktem2018bilingual} as the primary source of training/testing data for the task, and while this corpus is hand-aligned, it is too small (7K samples) to be used in the process of training deep neural networks. Instead, approaches turn to privately collected datasets \cite{singh2023looking}, or to datasets for simultaneous translation (ST) such as MuST-C \cite{di2019must} and MuST-Cinema \cite{karakanta2020must}. While ST datasets often have a large amount of source audio, they do not contain audio in the target domain, and cannot easily be used to evaluate prosody, lip-matching, timing, and spoken translation among other qualities. 

In this work, we introduce Anim-400K, a dataset of over 425K aligned dubbed clips designed from the ground up for synchronized multi-lingual tasks, such as automated dubbing. Anim-400K is over 40 times the size of existing aligned dubbed video datasets, and contains rich metadata support for numerous other challenging video tasks (\autoref{sec:tasks}). An outline of this paper is as follows: \autoref{sec:dataset} discusses the data collection process, the information collected, and compares Anim-400K to existing datasets, \autoref{sec:tasks} overviews some of the potential tasks that Anim-400K supports and \autoref{sec:limits}, discusses the limitations and ethics of the dataset.

\section{Dataset}
\label{sec:dataset}

\paragraph{Related Datasets:} As discussed in \autoref{sec:intro}, the availability of large-scale public research data has been a primary limiting factor in the development of dubbing methods \cite{agrawal2023findings}. An overview of related datasets is given in \autoref{tab:info}.  The only publicly available dataset designed explicitly for dubbing is the Heroes corpus \cite{oktem2018bilingual}, which contains 7,000 aligned clips translating from English (EN) to Spanish (ES). Unfortunately, the Heroes corpus is often too small to use for training simultaneous translation and dubbing models. Also too small for training models is the IWSLT 2023 test set, which contains only 200 clips collected in a constrained environment for dubbing from German (DE) to English (EN). Thus, most models turn to simultaneous translation (ST) datasets such as MSLT \cite{federmann2016microsoft} and MuST-C \cite{di2019must} for training. These datasets, while large, do not contain source video or target audio, and only contain text-translations of the data. Further, it is well known \cite{chafe1987relation} that spoken distributions of text differ from written distributions, and even more limiting, such translations do not need to conform to key dubbing metrics including prosody, isochrony, and timing. MuST-Cinema \cite{karakanta2020must} lies between ST and full dubbing, where the source video is provided, but the output still relies on translated subtitles instead of true dubbed audio. 

It is clear that a new large-scale dataset is required to fill the training gap between ST datasets and high-quality manually aligned datasets such as the Heroes and IWSLT corpuses. In this work, we focus on introducing this middle ground: a large-scale fully aligned dataset of audio segments containing true dubbed audio distributions.

\paragraph{Data Collection:} Anim-400K was sourced by scraping publicly available (ad-supported) dubbed anime videos from popular anime watching websites. At the time of scraping, none of the collected video was behind a paywall, or required any form of login to collect. We collected raw episodes in 1920x1080 resolution, 48KHz audio, with both Japanese and English audio tracks. We also collected the English subtitles for the Japanese language track. This collection process gives us unaligned dubs, as well as weakly aligned subtitles. In addition to collecting the visual information, we join metadata from a popular source for anime video metadata, and merged it with the collected video data. This enriches the collected data, and provides support for several additional tasks, which we describe in \autoref{sec:tasks}. An overview of the data is provided in \autoref{tab:info}.

\paragraph{Annotation:} A weakness of prior approaches \cite{oktem2017automatic, oktem2018bilingual} for collecting dubbed data is that they rely on a bottom-up approach for aligning audio clips, where individual words and segments are aligned using movie scripts, subtitles and other information. This leads to segments that match well with the audio, but are not necessarily fully aligned. Our approach, on the other hand, takes a top-down approach to extracting aligned segments, by ensuring that all segments are always aligned, but for noise (both ASR noise, and speaker noise) in the segment. This approach is additionally beneficial (or detrimental) in that it allows the model to capture unique performance content which may not be available in transcripts such as non-speech utterances.

\begin{table}[t]
    \small
    \centering
    \begin{tabularx}{\linewidth}{rX}
        \toprule
         \multicolumn{2}{c}{\textbf{Season/Show Information}} \\
         \midrule
         genres & The show genre (\hyperref[para:genre]{subsection 3.3}) \\
         themes & Themes in the show (\hyperref[para:genre]{subsection 3.3}) \\
         scores & User ratings (\hyperref[sec:vqa]{subsection 3.4}) \\
         characters & Character bios, pictures (\hyperref[sec:char]{subsection 3.2}) \\
         synopsis & Short show description (\hyperref[sec:vd]{subsection 3.1}) \\
         source info & Dates, Producers, Licensors, Studios etc \\
         \midrule
         \multicolumn{2}{c}{\textbf{Episode Information}} \\
        \midrule
         synopsis & Short episode description (\hyperref[sec:vd]{subsection 3.1}) \\
         scores & Use ratings (\hyperref[sec:vqa]{subsection 3.4}) \\
         subtitles & EN subtitles for JP audio (\hyperref[sec:st]{subsection 3.5}) \\
         \midrule
         \multicolumn{2}{c}{\textbf{Segment Information}} \\
         \midrule
         timing & Start/end times (EN/JP) \\
         speakers & Episode-specific IDs for contained speakers \\
         ASR & Aligned ASR transcript (EN/JP) \\
         \bottomrule
    \end{tabularx}
    \caption{Overview of the information contained in Anim-400K at the season/show, episode, and segment levels.}
    \label{tab:info}
    \vspace{-1em}
\end{table}

\noindent\textit{Aligned Clip Extraction:} To extract aligned clips from the raw video, we first use AWS Transcribe to create ASR transcripts of the spoken audio in both the Japanese (JP) and English (EN) versions of the episodes. Because the video is the same for each audio track, we know that the videos are globally temporally aligned. Thus, to generate local clips alignments, for each segment in the EN ASR transcript, we recursively merge the segments with other ASR segments (in either EN or JP) that have either overlapping endpoints, or endpoints differing by up to 125ms (which we found empirically to generate high quality segments). This process is repeated until no additional segments are added. For each clip, release the video, the timing (start/end times), and the ASR for both JP and EN, as well as any EN subtitles for JP audio that overlaps with the given clip.

\noindent\textit{Speaker Annotation:} In order to understand the content of each clip, we additionally use an off-the-shelf speaker diarization method, PyAnnote \cite{Bredin2021}, at an episode level to label speakers for each clip (made available in the dataset). In practice, we found that of the 437K clips in the Anim-400K dataset, 323K were judged to be single-speaker clips, while 114K were multi-speaker clips. We have marked these clips in the dataset, and these clips provide a challenging test for dubbing methods which must correctly isolate and reproduce several concurrent speakers, something no current system is capable of handling.  

\noindent\textit{Mixing and Cleaning:} To develop end-to-end dubbing libraries, it is often the case that generated text to speech audio will need to be mixed with a clean audio track to generate the final audio. In addition to the EN and JP audio tracks, we make available a further ``backing'' audio track, generated by running source separation tools against the JP audio \cite{spleeter2020}. This track, while sometimes noisy, generally provides a good baseline for new dubbing methods. We additionally further provide a mixing ratio for each clip: the ratio at which normalized audio should be mixed with the normalized backing track to closest approximate the overall audio mix, to avoid situations where the mixed TTS is much louder or softer than the related video. 

\noindent\textit{Baselines:}\label{sec:baselines} In addition to collecting the dataset, we also aim to allow for repeatable and robust evaluation of automated dubbing methods on the test partition of the dataset. While many methods use ``Mean Opinion Scoring (MOS)'' scores to evaluate their approaches, these ratings are well known to be dependent on a wide range of user-dependent factors \cite{schinkel2012does}. Instead, we recommend the use of MUSHRA (MUlti Stimulus test with Hidden Reference and Anchor) \cite{series2014method} to evaluate automated dubbing approaches on the Anim-400K dataset. MUSHRA involves presenting the listener with a specified quantity of test samples, a concealed variation of the reference, and one or more anchor points. To enable consistent MUSHRA evaluation, we provide two anchor tracks: the gold standard audio collected from the EN dub, and a baseline automatically generated dub, created from a simple pipeline. 

To generate the baseline dubbing tracks, we first split the audio into vocals and accompaniment using Spleeter \cite{spleeter2020}. We performed speaker diarisation to split all the multi-speaker Japanese clips into single-speaker segments to allow for better performance during the TTS using PyAnnote \cite{Bredin2021}. Afterward, we transcribed and translated each of the solo Japanese speaker segments to get the English text for the TTS using Whisper \cite{radford2022robust}. Finally, we performed TTS with the single-speaker vocal segment as the reference and the translated transcription as the text using YourTTS \cite{casanova2022yourtts} and recombine these vocal segments with the accompaniment audio.

\begin{table}[t]
    \footnotesize
    \centering
    \begin{tabularx}{\linewidth}{Xcccc} %
    \toprule
    \textbf{Dataset} & \textbf{Sentences} & \textbf{\makecell{Words/\\Sentence}} & \textbf{\makecell{Words/\\Clip}} & \textbf{\makecell{Sentences/\\Clip}}\\
    \midrule
    {Heroes (ES)} & 10K & 5.11 & 6.92 & 1.35\\
    {Heroes (EN)} & 10K & 5.64 & 7.99 & 1.41\\
    {Anim-400K (JP)} & 1.69M & 3.09 & 11.97 & 3.88\\
    {Anim-400K (EN)} & 1.20M & 5.80 & 15.96 & 2.75\\
    
     \bottomrule
    \end{tabularx}
    \caption{Overview of some differences in natural language distribution between the Heroes \cite{oktem2018bilingual} and Anim-400K datasets. }
    \label{tab:datastats}
\end{table}

\section{Supported Secondary Tasks}\label{sec:tasks}

In this section, we outline additional tasks supported by the Anim-400k dataset due to its robust metadata, beyond its primary purpose of end-to-end video dubbing. %

\paragraph{(3.1) Video Summarization/Teaser Generation:}\label{sec:vd} Recently, there has been significant scientific interest in summarizing and describing video as natural language descriptions of video have the potential to aid in accessibility, content understanding and generation, recommendation algorithms and information retrieval domains (among others) \cite{meena2023review}. Unfortunately, for long-form videos ($>30s$), data for such summarization tools is largely unavailable. To help remedy this, in addition to the aligned video clips, Anim-400K contains 3.5K human-generated short ($62.85 \pm 61.99$ word) teaser summaries of selected episodes, designed to describe the contents of the video to a potential watcher, and entice them to watch the video. While this data may not be enough to allow for training summarization models, it can support the evaluation of video summarization and teaser generation tools. 

\paragraph{(3.2) Character Identification \& Description: }\label{sec:char} Understanding, locating, and naming characters within larger properties is a challenging task, for which data support is generally lacking. These tasks can often form the backbone of complex visual description, search, and analysis systems. In order to support tasks such as character re-identification \cite{kurt2017image} and character description \cite{gan2017stylenet}, we additionally collect short descriptions (on average $109.77 \pm 142.89$ words) for 1828 characters across the 190 represented shows, as well as 7516 still images of each of these characters. This augmentation to Anim-400K aims to provide scholars with valuable resources to advance character-related research and applications, contributing to the broader field of multimedia analysis.

\begin{figure}[t]
    \centering
    \includegraphics[width=\linewidth,trim={10cm 0 7cm 0},clip]{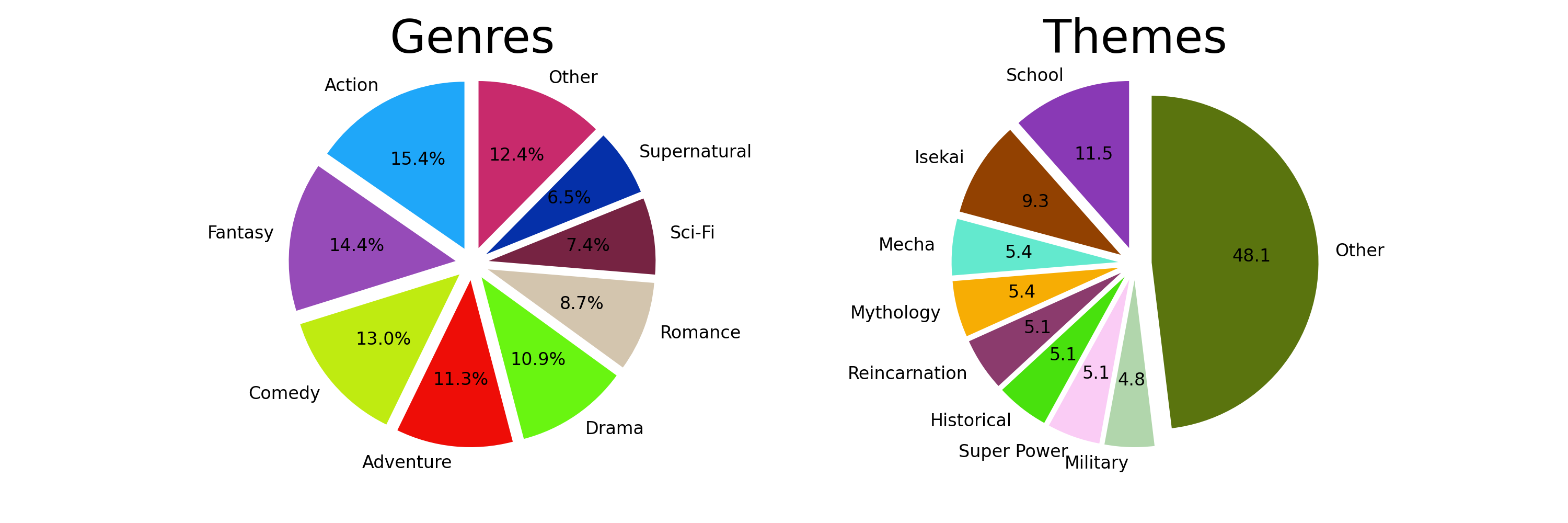}
    \caption{Genres and themes present in the Anim-400K dataset.}
    \label{fig:genres}
\end{figure}

\paragraph{(3.3) Genre/Theme/Style Identification:}\label{para:genre} Understanding the genres, themes, and styles present in animated video can have several applications, including content recommendation, audience targeting and content analysis among others. To support research in these domains, in addition to collecting the shows themselves, in Anim-400K, each show is labeled with at least one of 18 genres, and can contains up to 44 themes. The distributions across the most common themes are shown in \autoref{fig:genres}. There are an average of $2.84 \pm 1.29$ genres, and an average of $1.64 \pm 0.99$ themes per show. In addition to the simple classification tasks enabled by labeling with genre and themes, Anim-400K can support both the problem of art style classification, the process of determining if two images are from the same anime/series/studio, and art style transfer, the process of transferring images between styles, both of which have been well studied in prior work \cite{li2022challenging,li2023parsing}. Individual frames in Anim-400K extracted at a rate of 1FPS provides 2.3M  images  across the 190 properties in the dataset.

\paragraph{(3.4) Video Quality Analysis:}\label{sec:vqa} Video quality assessment models have garnered substantial attention, serving as a crucial toolkit utilized by the streaming and social media sectors \cite{tu2021rapique}. In today's media landscape, where content creators and platforms strive to capture and retain audiences, understanding the factors that contribute to overall property quality is paramount. To help enable research into video quality assessment, Anim-400K collects several metrics for quality at both the show and episode level. At the show level, we collect three measures of show quality: a numeric rating average across the users, the number of ``members" a show has (people who are following updates to the show), and the number of ``favorites" a show has (the number of people who have marked the show as a favorite). At the episode level, we collect the responses to user polls running shortly after each episode has aired, where users (on average $284.13 \pm 490.41$) rate the episode of the show on a 1-5 scale (with votes averaging a $4.23 \pm 0.65$). 

\paragraph{(3.5) Simultaneous Translation:}\label{sec:st} Simultaneous translation (ST) is often a sub-component of many dubbing systems, and consists of translating spoken audio into a text version of that audio in another language. Anim-400K further contains collected EN subtitles overlapping each collected audio clip. This data, similar in format to MuST-Cinema \cite{karakanta2020must}, allows for ST task support on Japanese, and Anim-400K is a relatively large dataset on a non-latin based language, making it a strong complement to any latin-based dataset such as MuST-C \cite{di2019must} when pre-training for ST or ASR tasks.

\section{Limitations \& Ethics}
\label{sec:limits}

The introduction of Anim-400K, while a substantial advancement, comes with notable ethical considerations and limitations. Firstly, there is a potential for data bias and a lack of representativeness, which may lead to skewed preferences or cultural insensitivity in the models trained on the dataset. This bias could result from the dataset not fully capturing the diversity of themes, genres, and cultural nuances present in the anime industry. In addition, because the dataset is limited to animated content, it likely will not transfer well to live-action media. Moreover, concerns about translation quality arise as automated dubbing relies on machine translation and voice synthesis technologies, which may not consistently meet high standards set by human translators and dubbing teams.

In addition to data bias limitations, it is important to recognize ethical considerations when using the dataset. Cultural sensitivity is paramount, as anime often includes culturally specific elements and references. Automatic dubbing systems must prioritize cultural competence and respect for the source material's context. Additionally, voice synthesis technologies may not fully replicate the nuances of human voice acting, potentially impacting the authenticity of dubbing and raising concerns about the replacement of human voice actors. Consent, copyright compliance, and user privacy are crucial aspects to consider when using the dataset for dubbing applications.

To address these limitations and ethical challenges, ongoing monitoring, evaluation, and refinement of automatic dubbing systems are essential. Collaborative efforts between researchers, developers, and the community can ensure responsible and respectful use of the dataset, enhancing the digital video viewing experience while upholding cultural sensitivity, translation quality, and ethical standards.

\section{Conclusion}

In conclusion, the Anim-400K dataset offers a substantial resource for automated dubbing with over 425K aligned dubbed clips, significantly surpassing existing datasets in size, and the dataset's rich metadata extends its usability to various video-related tasks beyond dubbing. While it holds great promise for improving accessibility and engagement, it's important to acknowledge the ethical and practical limitations associated with such large-scale datasets, and as we explore the potential of end-to-end dubbing and related fields, responsible development and ethical considerations should guide our efforts to ensure inclusivity and respect for cultural boundaries.
\clearpage
\section{References}
\vspace{-1em}
\begingroup
  \def\section*#1{}
  \small
  \setlength{\bibsep}{4pt}
  \bibliographystyle{IEEEtranN}
  \bibliography{refs}
\endgroup

\end{document}